\documentclass[preprint,review,12pt]{elsarticle}%
\usepackage{amsmath}
\usepackage{geometry}
\usepackage{amssymb}
\usepackage{amsthm}
\usepackage{amsfonts}
\usepackage{amsmath}
\usepackage{graphicx}
\usepackage{setspace}
\usepackage{fancyhdr}
\usepackage{fancyvrb}
\usepackage{lineno}
\usepackage[numbers]{natbib}
\usepackage{multirow}%
\providecommand{\U}[1]{\protect\rule{.1in}{.1in}}
\allowdisplaybreaks[1]
\renewcommand{\mathbf}[1]{\mbox{\boldmath $#1$ \unboldmath}  \!\!}
\journal{Journal of Non-Newtonian Fluid Mechanics}
\begin{document}

\begin{frontmatter}
\title{A simple and efficient numerical procedure to compute the inverse Langevin
function with high accuracy}
\author{Jos\'{e} Mar\'{i}a Ben\'{i}tez}
\author{Francisco Javier Mont\'{a}ns\corref{cor1}}
\address{Escuela T\'{e}cnica Superior de Ingenier\'{\i}a Aeron\'{a}utica y del Espacio\\Universidad Polit\'{e}cnica de Madrid\\
Plaza Cardenal Cisneros, 3, 28040-Madrid, Spain}
\cortext[cor1]{Corresponding author. E-mail: fco.montans@upm.es}
\begin{abstract}
The inverse Langevin function is a fundamental part of the statistical chain
models used to describe the behavior of polymeric-like materials, appearing
also in other fields such as magnetism, molecular dynamics and even
biomechanics. In the last four years, several approximants of the inverse
Langevin function have been proposed. In most of them, optimization techniques are used to reduce the relative error of previously published approximants to reach orders of magnitude of $O\left(  10^{-3}\%-10^{-2}\%\right)  $. In this
paper a new simple and efficient numerical approach to calculate the inverse
Langevin function is proposed. Its main feature is the  reduction of the relative errors in all the domain $x=[0,1)$ to near machine precision, maintaining function evaluation CPU times  similar to those of the most efficient approximants.  The method consists in the discretization of the Langevin function, the calculation of the inverse of these discretization
points and their interpolation by cubic splines. In order to reproduce the asymptotic behavior of the inverse Langevin function, a rational function is considered only in the asymptotic zone keeping  $\mathcal{C}^{1}$  continuity with the cubic splines. We include customizable Matlab codes to create the spline coefficients, to evaluate the function, and to compare accuracy and efficiency with other published proposals.\end{abstract}
\end{frontmatter}

\section{Introduction}

The inverse Langevin function is frequently used in fields like polymer
science \cite{ammar2016effect,kroeger2015simple,underhill2007accuracy}, rubber hyperelasticity \cite{arruda1993three,miehe2004micro} and biomechanics \cite{benitez2017mechanical}. In these
contexts, a single molecular chain can be modeled as a freely-joined chain
(FJC), which is composed of $n$ linked rigid segments of equal length $l$,
randomly coiled, so that $L=nl$ is the contour length of the chain. Its
configuration is defined by the end-to-end distance or tie points distance,
$r$, that is the length between the ends of the chain. When $r=0$, both points
are coincident and the entropy is maximum, whereas if the chain is
cross-linked, the entropy decreases since the number of possible
configurations that the chain can take also decreases \cite{bergstrom2015mechanics,treloar1975physics}. The entropy of a molecule network, $S$, composed
by $N$ chains per reference volume, can be defined in terms of the available
configurations, which are mathematically given by a probability distribution
function (PDF) $p\left(  r\right)  $%
\begin{equation}
S=c+Nk_B\ln\left(  p\left(  r\right)  \right)
\end{equation}
where $c$ is a constant and $k_B$ is the Boltzmann constant.

When a single chain, $N=1$, develops finite strain, Gaussian PDFs are no
longer valid since they do not account for the limit of the chain
extensibility when $r\rightarrow L=nl$. Thus, a PDF based on the inverse
Langevin function is used so that a relationship between the force applied in
a single chain and its extensibility can be obtained through%

\begin{equation}
f=-T\frac{\partial S}{\partial r}=\frac{k_B}{l}T\mathcal{L}^{-1}\left(  \frac
{r}{nl}\right)
\end{equation}
where $T$ is the absolute temperature and $\mathcal{L}^{-1}$ is the inverse
Langevin function \cite{treloar1975physics}. The Langevin function is defined as%

\begin{equation}
x=\mathcal{L}\left(  y\right)  =\coth\left(  y\right)  -\frac{1}{y}
\label{Lang}%
\end{equation}

There is no explicit form of its inverse, $\mathcal{L}^{-1}\left(  x\right)
$, and several approximants have been proposed. The first one was proposed by
Kuhn and Gr\"{u}n \cite{kuhn1942beziehungen} and consists in a Taylor
expansion series around $x=0$. Although, as expected, approximants based on Taylor expansion
series show a good accuracy around the fixed point $x=0$, they have divergence problems in the
vicinity of $x=1$, when the chain starts to be fully stretched, so
$r\rightarrow L=nl$. Recent studies have proven that with $115$ series terms,
a good accuracy is found in the interval $\left[  0,0.95\right] $ and that
above $500$ terms the convergence radius remains unchanged
\cite{itskov2012taylor,jedynak2017new}.
It is obvious that the computational evaluation of such number of terms is exceedingly expensive.

To account for the singularity of the inverse Lanvegin function at $x=1$, several
Pad\'{e} approximants have been proposed. These approximants are rational
functions denoted by $\left[  m/n\right]  $, where $m$ and $n$ are the degree
of the polynomials corresponding to the numerator and the denominator,
respectively. They account for the singularity at $x=1$ in terms of $1/\left(
1-x^{i}\right)  ^{j}$, being $i$ and $j$ natural numbers excluding the zero.
Examples of this kind of approximant are the proposals of Warner
\cite{warner1972kinetic}, Cohen's rounded Pad\'{e} proposal \cite{cohen1991pade} and Puso
\cite{puso2003mechanistic}. These approximants are usually known as
single-point Pad\'{e} approximants, since the polynomial coefficients are
determined with the information of the function $\mathcal{L}^{-1}\left(
y\right)  $ and its derivatives that is given by Taylor expansion series around $x=0$.
When these coefficients are calculated by evaluating the function in some
points within the domain $\left[  0,1\right)  $ the approximants are called
multi-point Pad\'{e} approximants. Examples of multi-point Pad\'{e}
approximants have been proposed by Jedynak \cite{jedynak2015approximation} and
Darabi and Itskov \cite{darabi2015simple}. Kr\"{o}ger  \cite{kroeger2015simple} improved some of the
aforementioned single- and multi-point Pad\'{e} approximants considering the
exact asymptotic, symmetry and integral behavior of the inverse Langevin
function.

In the last years, a new trend of improving some of the existing approximants
by optimization methods has emerged. Such is the case of the error-corrected
approximants developed by Nguessong et al. \cite{nguessong2014new}. They
improved the accuracy of some of the above mentioned multi-point Pad\'{e}
approximants calculating the error functions corresponding to the approximant
and subtracting both. To do so, Neguessong et al. \cite{nguessong2014new} developed an optimization of
the error function parameters by a least squares minimization. Following this
idea, Marchi and Arruda  \cite{marchi2015error} proposed not only the optimization of these parameters
but also the polynomial coefficients of the approximant to minimize its
relative error. Recently, Petrosyan
\cite{petrosyan2017improved} deduced a function
accounting for the asymptotic behavior of the inverse Langevin function and
minimized its absolute error with a sine and a quadratic function.

Also recently, Jedynak \cite{jedynak2017new} improved the approximants given by Kr\"{o}ger
\cite{kroeger2015simple} and proposed a new one applying the minimax
approximation theory. The approximation theory is a method to determine the
degree of the polynomial or rational approximation that minimizes the error of
a certain function. Jedynak minimized the relative error of the inverse
Langevin function increasing the complexity of the Kr\"{o}ger   interpolant to $11$, that is to say with a rational function of $[9/2]$
degree. The polynomial coefficients were determined solving a system of
nonlinear equations by means of Remez's algorithm.

An approximant not based on Taylor series or Pad\'{e} functions was proposed
by Bergstr\"{o}m. It is  a piecewise function resulting from dividing the
domain in two subdomains to account for the asymptotic behavior of $\mathcal{L}%
^{-1}$, Ref. \cite{bergstrom1999large}. Despite having good accuracy,
this approximant has some drawbacks that make its application to physical
problems difficult, see Refs. \cite{jedynak2015approximation} and \cite{marchi2015error}.

The fact that most of the above-mentioned papers have been published in the
last four years, from $2014$ to $2017$, shows that the
calculation of the inverse Langevin function is of much current interest and that a
satisfactory, computationally efficient and accurate solution has not been reached yet \cite{darabi2015simple,jedynak2015approximation,kroeger2015simple,petrosyan2017improved,jedynak2017new,marchi2015error}.
These works have been mainly focussed on improving the accuracy of previous
researches. To summarize and to compare the accuracy of the main approximants
proposed, their maximum relative error, ${\Large\varepsilon}_{r}$, is shown
in Table \ref{rel-err-1}. According to these data, a lot of effort has been
invested to reduce the maximum relative error of the rounded Cohen approximant in
only one order of magnitude, $O\left(  10^{-1}\%\right)  $ in the case of
Kr\"{o}ger or Petrosyan. The order of magnitude of the error has not  improved
substantially with the error minimizing techniques. With this optimization
methods the order of magnitude of the relative error is $O\left(
10^{-2}-10^{-3}\%\right)  $ \cite{jedynak2017new,nguessong2014new,marchi2015error}.

\begin{table}
\label{rel-err-1}%
\centering
\begin{tabular}
[c]{ll}\hline
Approximant & ${\Large \varepsilon}_{r}$ $\left(  \%\right)  $\\\hline
Cohen (1991) \cite{cohen1991pade} & $4.94$\\
Kr\"{o}ger (2015)\cite{kroeger2015simple} & $2.75\cdot10^{-1}$\\
Petrosyan (2017) \cite{petrosyan2017improved} & $1.79\cdot10^{-1}$\\
Nguesson et al. (2014) \cite{nguessong2014new} & $4.65\cdot10^{-2}$\\
Jedynak (2017) \cite{jedynak2017new} & $7.69\cdot10^{-2}$\\
Marchi and Arruda (2015) \cite{marchi2015error} & $4.37\cdot10^{-3}$\\
\hline%
\end{tabular}
\caption{Maximum relative error of the main approximants.}%
\end{table}

The accuracy of the inverse Langevin function is not a minor issue since it has a relevant influence in the
results obtained in computational simulations. Indeed, the inverse Langevin
function is used in many models implemented in finite element codes
\cite{ammar2016effect,bischoff2000finite,bischoff2002finite,bischoff2002microstructurally,fleischhauer2012constitutive}, including commercial finite element programs such as ABAQUS or ADINA. During simulations, the inverse Langevin function is typically evaluated  millions of times, so iterative methods are avoided and explicit approximants are preferred.
Recently, Ammar has shown the important influence of the approximants accuracy in the
results obtained in simple finite element simulations in the framework of
the dilute polymer kinetic theory \cite{ammar2016effect}. When complex
calculations are performed, the approximant accuracy can be a critical factor
to obtain reliable results. Therefore, an approximation of the inverse
Langevin function with a simple and computationally efficient implementation
in a finite element code, and with a suitable accuracy, is needed to ensure
adequate efficient computational predictions. 

In this paper we present a simple computationally-oriented technique to calculate the inverse
Langevin function that reduces the maximum relative error close to the one that a
computer can obtain. In this proposal, the inverse Langevin function is obtained by means of a  cubic spline representation of the function. However, to reproduce the asymptote at $x=1$, the discretization close to the asymptotic zone is replaced by a $[1/2]$ rational function that fulfills $\mathcal{C}^{1}$ continuity conditions with cubic splines, so constitutive tangents keep continuity. Thus, the approximation to the inverse Langevin function
consists in a series of piecewise polynomials and a rational function in the vicinity of the asymptote. The derivatives and integrals are also immediate (explicit) and continuous. We note
that although this method requires the calculation of the spline coefficients, these
coefficients are obtained just once. In fact, the coefficients can be previously stored and embedded in the code at program compilation time. The price to pay  for high accuracy and efficiency is an increased storage. However, the storage needs are negligible for current digital devices. For example, if the Langevin function discretization
is performed in $10,001$ points (a number of points that gives high accuracy as shown below), $10,000$ cubic splines must be calculated. Considering that every polynomial has
four coefficients, $40,000$ coefficients have to be stored in the computer
memory. Taking into account that a double precision number is stored in $8$
bytes, $320$kB are necessary to store all the spline coefficients, which is at least four orders of magnitude less than the typical RAM memory (e.g. $6$GB) of an economic laptop computer or a graphic card (GPU), and also several orders of magnitude less than the typical memory needed to solve an industry problem.
Near machine precision is obtained with $100,000$ spline pieces which takes just about $3$MB of memory.
Remarkably, function evaluation times are almost independent of the desired, and obtained, accuracy.

Matlab codes are given in the appendices and commented in the text below. Although
the method is explained using Matlab, its implementation in any computer language is
straightforward.
 
\section{Determination of the inverse Langevin function by cubic splines
interpolation}
\subsection{Splines}
Piecewise cubic splines are cubic interpolating polynomials which have minimum curvature and preserve continuity of derivatives between contiguous segments up to second order. Some background on splines may be found, for example in Reference
\cite{de2005spline}. Some applications in the context of  hyperelasticity in polymers and biological tissues may be found, for example, in  \cite{latorre2015material,latorre2017determination,latorre2013extension,crespo2017wypiwyg,romero2017determination}.

Splines may interpolate data within some interval. Outside the interval, end conditions determine the behavior. Each polynomial is valid within some subinterval, and its expression may be written as (Matlab names this form as \emph{piecewise polynomial form}; ``pp-form'')%

\begin{equation}
y\left(  \hat{x}\right)  =%
{\displaystyle\sum_{k=1}^{4}}
C_{k}\left(  s\right)  \left(  \hat{x}-x_{i}\right)  ^{4-k}%
\end{equation}
where $x_{i}$ is the lower limit of subinterval $s$ and $C_{k}\left(  s\right)$ are the coefficients of the spline. The determination of these coefficients is highly efficient (with times of the order of the number of interpolation points) because matrices are tri-diagonal and the system of equations may be solved using the efficient Thomas algorithm; see for example \cite{de2005spline}. Matlab has a splines toolbox which we use in this paper to develop the demonstrative codes given in the appendices.    
\subsection{Procedure to determine the interpolation coefficients}
\begin{figure}
[ptb]
\begin{center}
\includegraphics[width=1.0\textwidth]{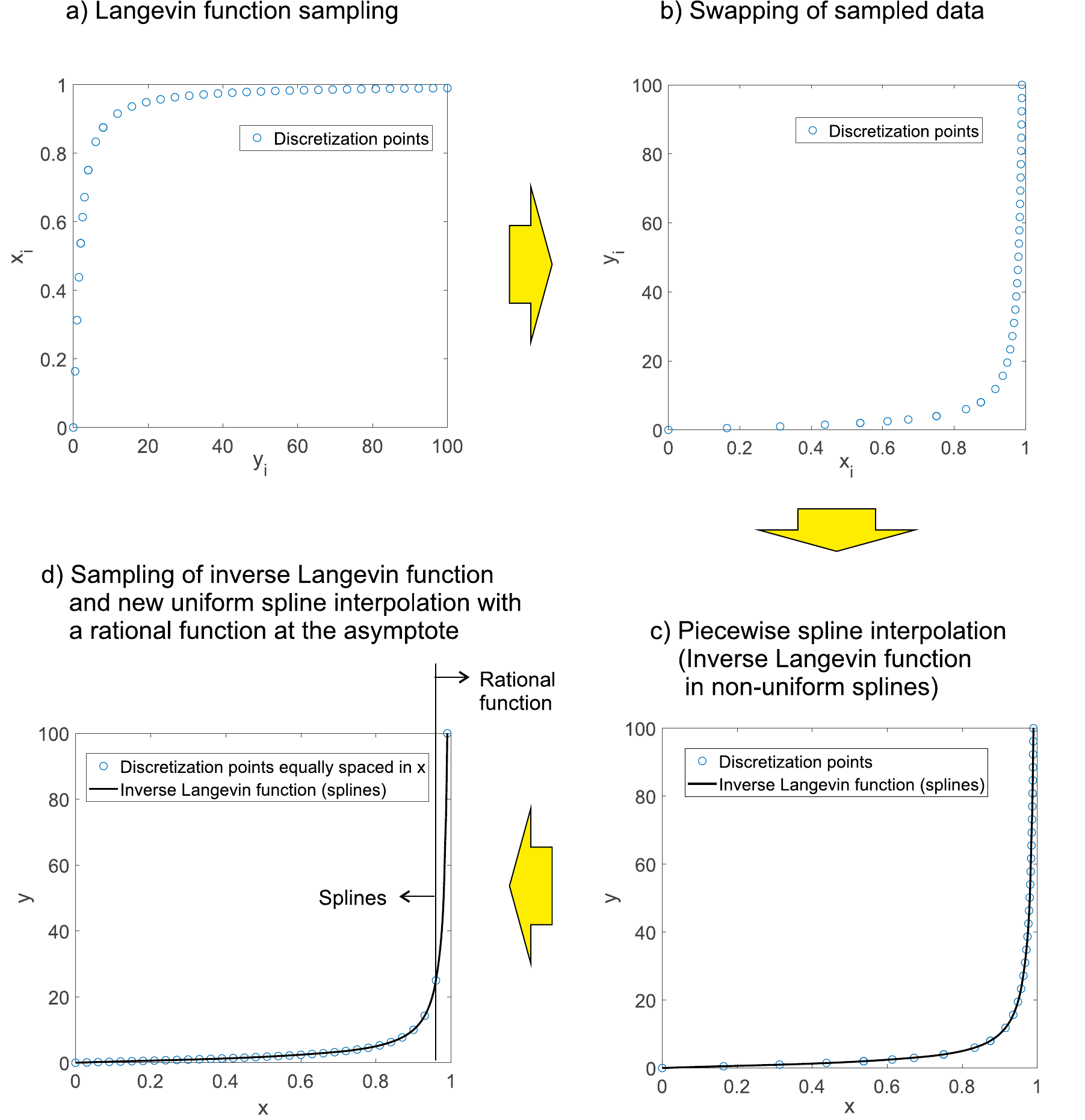}%
\caption{Description of the method to approximate the inverse Langevin function. a) Sampling of the Langevin function $\{x_i,y_i=\mathcal{L}(x_i)\}$. b) Swapping of the data $\{y_i=\mathcal{L}(x_i),x_i\}$. c) Piecewise (nonuniform) spline interpolation of $\{y_i,x_i=\mathcal{L}^{-1}(x_i)\}\rightarrow x(y) \simeq \mathcal{L}^{-1}(y)$. d) Resampling of $x(y) \simeq \mathcal{L}^{-1}(y)$ in uniform intervals and new piecewise spline interpolation for efficient evaluation. Last interval is evaluated using a $\mathcal{C}^{1}$ compatible rational function instead of the spline.}%
\label{Figura_metodo.pdf}%
\end{center}
\end{figure}

The proposed method to calculate the inverse Langevin function has two custom parameters that control accuracy and storage needs. These parameters are the number of spline pieces $n$ (variable \tt nsp = npoints - 1 \rm in the code in the appendices) and the initial point $x_{r}$ for the appended rational function (variable  \tt xir \rm in the code) to accommodate the asymptotic behavior. The inverse Langevin correspondence $y_r$ (variable \tt yir \rm in the code) of $x_r$ can be given beforehand or computed. Then, the procedure to compute the spline coefficients involves five
steps. These steps are shown graphically in Figure  \ref{Figura_metodo.pdf} and can be summarized as:

\begin{enumerate}
\item Sampling of the Langevin function  at locations: $\{  y_{i},x_{i}%
=\mathcal{L}\left(  y_{i}\right)  \} , i=1,...,2n+1  $, with $y$ taking values from $y=0$ to $y=2y_r$; Fig.  \ref{Figura_metodo.pdf}a.  The factor of two is included to avoid known spline border effects at the turning point of the Langevin function. 

\item Swapping of the abscissa and ordinate columns (this is a conceptual step, since it does not entail any actual operation; it is simply the interchange of columns in an ideal table): $\{ x_{i}%
=\mathcal{L}\left(  y_{i}\right)  ,y_{i}\}  $; Fig.  \ref{Figura_metodo.pdf}b.

\item First interpolation of $\{  x_{i},y_{i}\}  $, from $y=0$ to $y=2y_r$ with cubic splines using $2n$ segments. These splines have equispaced ordinates $y$ and are not efficient for the evaluation of the inverse Langevin function for a given $x$; see Fig.  \ref{Figura_metodo.pdf}c.
End conditions are not important here, so second derivatives are taken as vanishing.\item Determination of a new set of computationally efficient uniform splines, where the abscissae $x$ are equispaced and known end derivatives are enforced; i.e. $dy/dx=3$ at $x=0$ and $dy_r/dx_r=y_r^2/(1+y_r^2-y_r^2\coth^2y_r)$ at $(x_r,y_r)$. For given $x_j=(j-1)\Delta x$ with $j=1,...,n$, the approximate value $y_j$ is obtained from the previous spline; see Fig.  \ref{Figura_metodo.pdf}d.\item Attaching of the rational function $y=(ax+b)/(1-x^2)$ for values $x>x_r$.    The coefficients $a$ and $b$ are obtained from the continuity conditions $y(x_r)=y_r$ and $dy/dx=dy_r/dx_r$ as given above. This rational function accommodates the asymptote which cannot be captured by splines; see Fig.  \ref{Figura_metodo.pdf}d.
\end{enumerate}

The fully commented Matlab code in Appendix A reproduces the previous steps and saves the coefficients in a Matlab workspace file \tt InverseLangevinCoefs.mat \rm under  variable \tt Coefs \rm (the spline pp-form) for downstream loading and use. 
\subsection{Comments on custom parameters determining accuracy and storage needs}

\label{procedure}

The previous procedure is of course  not the only one  possible, and not even the most efficient; it is a simple implementation of the idea. In particular, we have used a uniform spacing of the first spline in the $y-$axis and a uniform spacing of the second spline in the $x-$axis. For the first spline, non-uniform spacing may provide better accuracy with substantially fewer interpolation points. For the second spline, non-uniform spacing with a formula for indirect addressing in the applicable interval may also substantially improve accuracy with fewer polynomials. However, we prefer here to keep simplicity in the exposition of ideas and in the related code.
   
In this context, for the specific procedure detailed above which code is given in Appendix A, the parameter $n$ controls the accuracy of the method. For a given value of $n$, there are some values of $x_r$ which give a good balance in the errors obtained from splines and the rational function. Some examples are given as comments in the code in the appendix. The main issue in improving the accuracy of the procedure is a good discretization of the turning zone of the Langevin function.   

As it will be seen below, with $n=10^5$, the above procedure reaches in practice machine precision in most of the domain, but storage needs in Matlab are  $3,361$kB (spline-break values are included in this memory for convenience, but can be eliminated). A lower number, $n=10^4$ uses just $365$kB and gives relative errors which are only between one and two orders of magnitude above machine precision. 

Regarding computational times, it is important to remark that the function evaluation times are virtually independent of the number of subintervals $n$. Hence in this regard, the maximum accuracy  $n=10^5$ should be used. Even the computational time of the spline coefficients for this large number is very low. However, we also remark that the computation of the spline coefficients should not be included in a final code; these are computed once when building a code, after a decision on the precision of the code is made. In Matlab, the coefficients may be loaded any time from a file. In fortran, the coefficients may be explicitly given, for example, in \tt data  \rm statements in the source code to be compiled, so they are readily available in the compiled executable (\tt .exe \rm) code. 

\section{Comparison with other approximants}

There is no known analytical expression for the inverse Langevin function.
It is
obvious that a numerical solution may be obtained from Newton-Raphson iterations or similar procedures. However, as mentioned in the Introduction Section, the  inverse Langevin function is typically evaluated millions of times in a computer code, so computational efficiency is important, and Newton iterations are exceedingly expensive for this task. Therefore several approximants have been recently proposed to improve the accuracy of the evaluations keeping the computational cost of such evaluations low.
In this section we compare  the accuracy and computational times of our procedure to those obtained by some recent proposals. We include in Appendix B the Matlab code with which we have performed these comparisons and created the figures. 
\subsection{ Load-type accuracy}
  
Assume that a load-type  value $y$ is given. The accuracy of the approximants to the inverse Langevin function is usually measured through the relative error in the load $y$ for a given $x$ as

\begin{equation}
\label{Rerror}
{\Large \varepsilon}_{r}{\Large =}\frac{\left\vert y-\mathcal{L}%
^{-1}\left(  \mathcal{L}\left(  y\right)  \right)  \right\vert }{y}\times 100
\end{equation}where $\mathcal{L}\left(  y\right) \equiv x  $ is the Langevin function (which analytical expression is known) and $\mathcal{L}%
^{-1}\left(  \mathcal{L}\left(  y\right)  \right) \equiv \mathcal{L}%
^{-1}\left( x  \right)$ is the approximation to the inverse Langevin function (i.e. the approximant we are evaluating). Frequently, maximum error values for the approximants are given. However, we consider that the error spectrum in the full domain gives a better picture because as seen below, achievable machine errors are also different in different parts of the domain. 

\begin{figure}
[ptb]
\begin{center}
\includegraphics[width=0.8\textwidth]{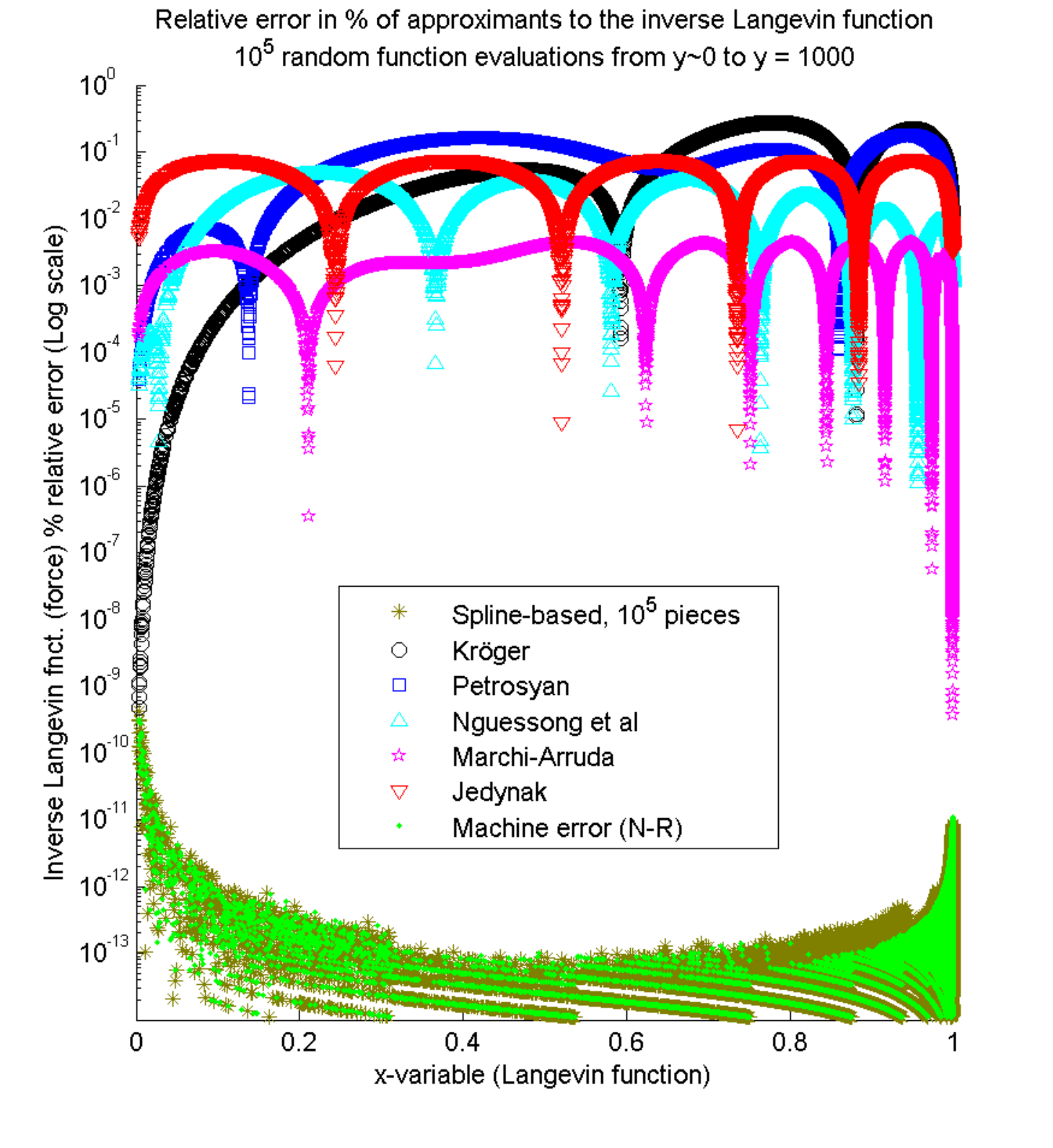}%
\caption{Comparison of the accuracy of the proposed method with that of some published approximants. Proposed method computed with $n=10^5$ and $x_r=0.98$. Compared approximants are those of Kr\"oger \cite{kroeger2015simple}, Petrosyan \cite{petrosyan2017improved}, Nguessong et al \cite{nguessong2014new}, Marchi and Arruda \cite{marchi2015error} and Jedynak \cite{jedynak2017new}. Machine error has been estimated by running Newton-Raphson (N-R) iterations until no further convergence is detected and using Kr\"oger's approximant as initial guess.}%
\label{Accuracy1e5pieces.pdf}%
\end{center}
\end{figure}

In Figure \ref{Accuracy1e5pieces.pdf} we show a comparison of the present proposal using $n=10^5$ and $x_r=0.98$ with the approximants of Kr\"oger \cite{kroeger2015simple}, Petrosyan \cite{petrosyan2017improved}, Nguessong et al \cite{nguessong2014new}, Marchi and Arruda \cite{marchi2015error}, and Jedynak \cite{jedynak2017new}. The formulae used for these approximants are given in the references and in the code in  Appendix B. In order to estimate the machine error, i.e., the maximum accuracy which could be obtained from an iterative procedure, we have performed Newton-Raphson iterations until no further convergence is detected. In order be in the attraction zone of the Newton procedure towards the solution, we have employed Kr\"oger's approximant as an initial guess. The mean number of iterations needed over the approximant has been $4.3$. Remarkably, in Figure \ref{Accuracy1e5pieces.pdf} it can be seen that the spline-based procedure reached almost that machine precision in all the domain. In fact, no relevant improvement can be obtained with the spline-based procedure usin¡g more subintervals. This accuracy is many orders of magnitude better than that obtained using any¡ other approximant: note that the figure is in logarithmic y-scale. 

\begin{figure}
[ptb]
\begin{center}
\includegraphics[width=0.8\textwidth]{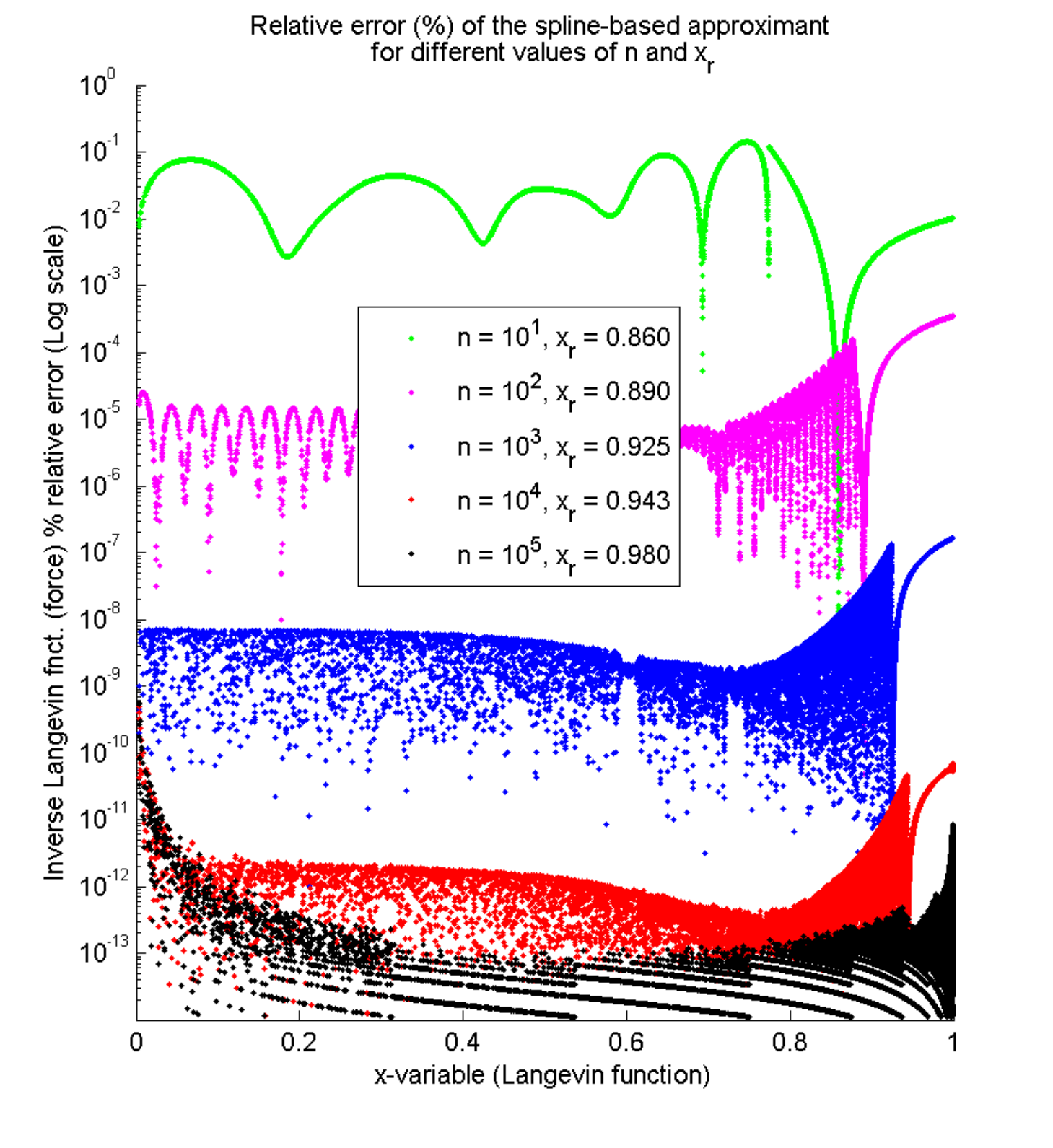}%
\caption{Comparison of the accuracy of the spline-based procedure for different values of the custom parameters $n$ and $x_r$. We note that $10^5$ random $y$ values have been generated and those same values have been used for all five sets in the Figure.}%
\label{Comparison_spline_different_n.pdf}%
\end{center}
\end{figure}

The influence of parameter $n$ in the accuracy of the spline-based approximant can be seen in Figure \ref{Comparison_spline_different_n.pdf}. In this figure we plot five possibilities given as comments in the code in Appendix A. It is seen that even with $n=10^4$ an excellent accuracy is obtained, in general only between one and two orders of magnitude  than machine precision. This number of subintervals may be a good compromise between excellent accuracy and limited amount of memory; hence $n=10^4$ and $x_r=0.943$ are our recommendation for $n$ and $x_r$. 
We finally note that it is observed that even with just $10$ subintervals a good accuracy is obtained.

\begin{figure}
[ptb]
\begin{center}
\includegraphics[width=0.8\textwidth]{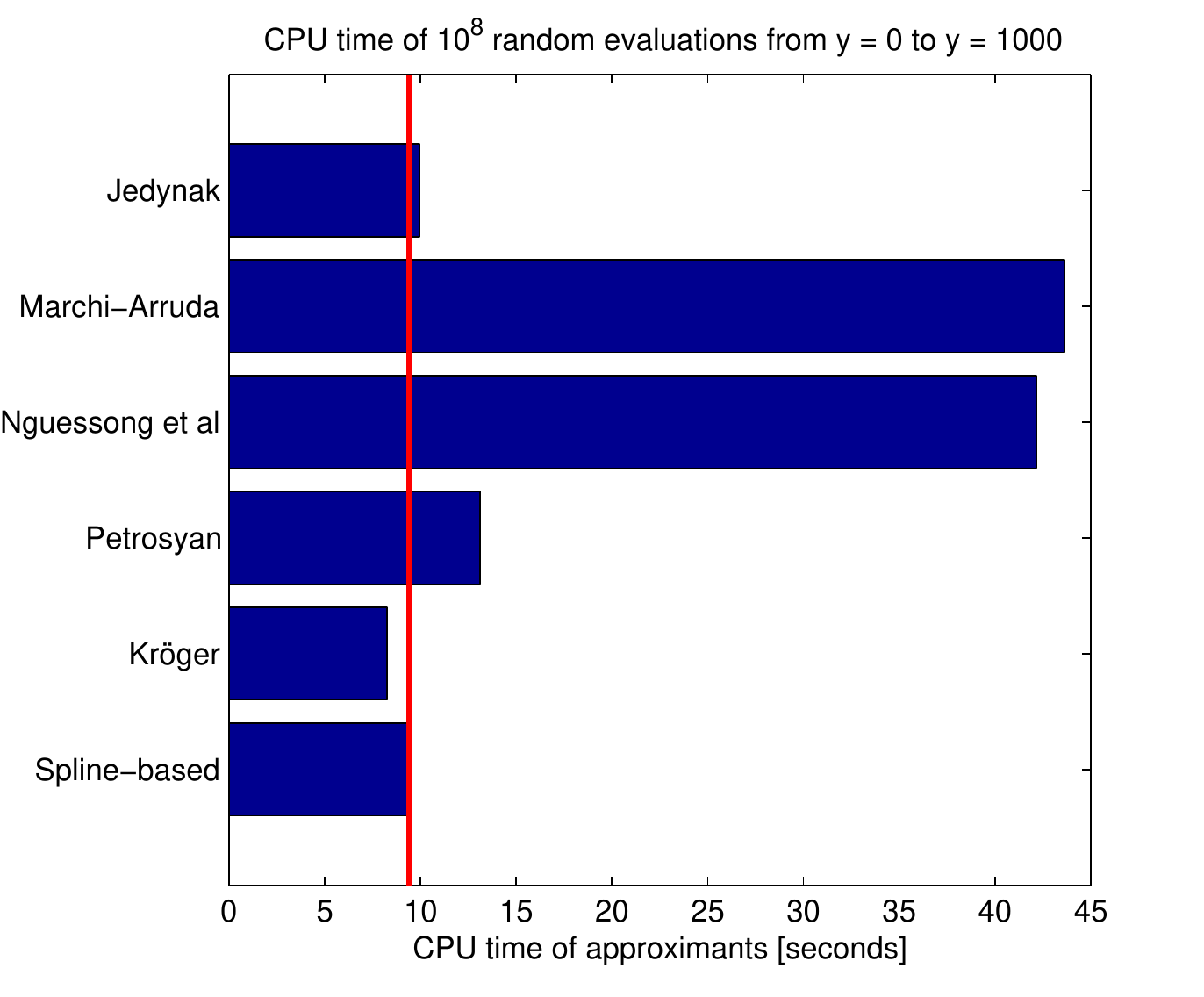}%
\caption{Comparison of the of the efficiency (CPU times in seconds) of the proposed method with that of some published approximants. Number of function evaluations is $10^8$ to reach stable CPU times.  CPU times obtained with Matlab R2011a (7.12.0.635) version in a Microsoft Surface Pro, Intel I5-6300U@2.4GHz 64-bit processor with Windows 10 Pro operative system. Proposed method computed with $n=10^5$ and $x_r$=0.98. Compared approximants are those of Kr\"oger \cite{kroeger2015simple}, Petrosyan \cite{petrosyan2017improved}, Nguessong et al \cite{nguessong2014new}, Marchi and Arruda \cite{marchi2015error} and Jedynak \cite{jedynak2017new}. For comparison, Newton-Raphson CPU time (not included in the figure) to reach machine precision departing from Kr\"oger's approximant has been of $5,430$ seconds (one hour and a half).}%
\label{CPU_vf_long_time2.pdf}%
\end{center}
\end{figure}

 \subsection{Computational efficiency}

CPU times depend on
different issues. An efficient implementation, precomputing repeated powers in formulae, may lower substantially the CPU time of one approximant. In the code given in appendix we coded approximants taking into account  efficiency. However, the tasks of the computer in the background, CPU temperature and overall load, memory and its speed, etc, also may strongly affect reported computational times. Hence, the CPU times given in Figure \ref{CPU_vf_long_time2.pdf} are just indicative of  relative efficiency among approximants. This plot is also generated by the code in the appendices. ¡However, to obtain more stable CPU times for a more meaningful comparison, we have employed for Figure \ref{CPU_vf_long_time2.pdf} as many as $10^8$ function evaluations. 

In Figure \ref{CPU_vf_long_time2.pdf} it can be observed that all approximants employ similar computational times, except the Nguessong et al \cite{nguessong2014new} and the Marchi and Arruda \cite{marchi2015error} approximants. The reason seems to be that the efficiency of power products with integer exponents is about four times better than those using real exponents, as it can easily be checked in Matlab. This is in line, even in magnitude, with the increased reported CPU times of these two approximants. In most of our simulations, Kr\"oger's approximant was the fastest one. Remarkably, all the cases reported in Figure \ref{Comparison_spline_different_n.pdf} present similar computational times, regardless of the values of $n$ and $x_r$: the CPU time of spline-based approximants is almost insensitive of the pursued accuracy. 

Finally, the importance of obtaining efficient approximants becomes apparent when analyzing computational times of the Newton-Raphson procedure. Even though we started the iterations from a good approximant as to be in the attraction zone of the solution and obtain second order convergence, it took about  $500$ times longer to reach machine precision than with the spline-based procedure; i.e. one hour and a half compared to just $10$ seconds! A single Newton-Raphson iteration took near $100$ times more than a function evaluation.

\subsection{Integral accuracy criteria}

Kr\"{o}ger and Jedynak proposed a second criteria to evaluate the accuracy of
the approximations of the inverse Langevin function \cite{kroeger2015simple,jedynak2017new}. This accuracy
measurement is given for the approximant capabilities to solve exactly the
following energy-related integral%
\begin{equation}
\frac{1}{2}\int_{0}^{\infty}\left[  1-\mathcal{L}\left(  y\right) \right]^2  dy 
=0.7606614015
\label{int}
\end{equation}

The integration of Eq. (\ref{int}) can be performed considering the inverse Langevin function

\begin{equation}
\frac
{1}{2}\int_{0}^{\infty}\left[  1-\mathcal{L}\left(  y\right) \right]^2  dy 
=\int_{0}^{1}\left(  1-x\right)  \mathcal{L}^{-1}\left(  x\right)  dx
\label{int2}
\end{equation}

\begin{figure}
[ptb]
\begin{center}
\includegraphics[width=0.8\textwidth]{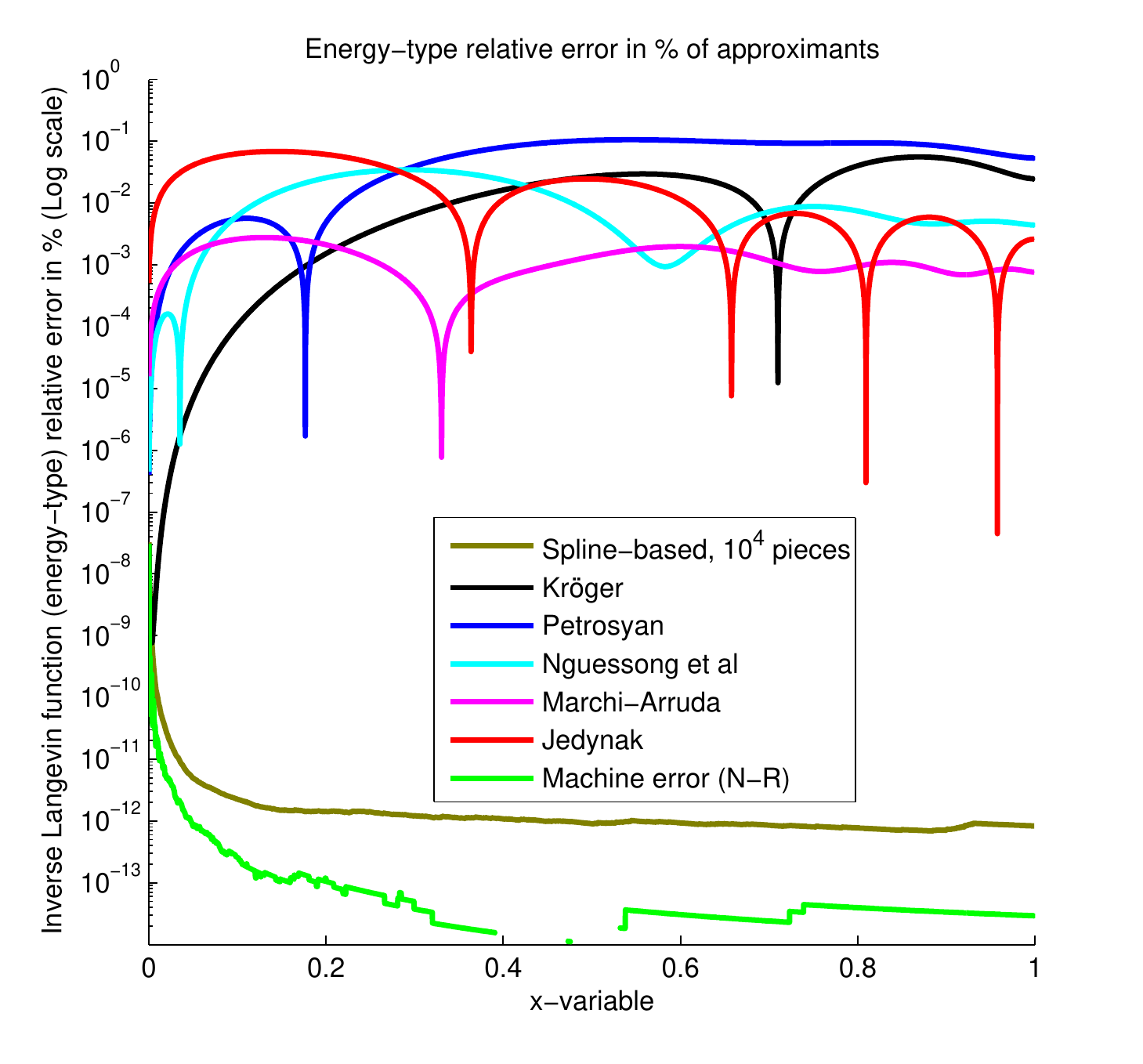}%
\caption{Comparison of the accuracy of the proposed method with that of some published approximants. Energy-type error. Proposed method computed with $n=10^4$ and $x_r=0.943$. Compared approximants are those of Kr\"oger \cite{kroeger2015simple}, Petrosyan \cite{petrosyan2017improved}, Nguessong et al \cite{nguessong2014new}, Marchi and Arruda \cite{marchi2015error} and Jedynak \cite{jedynak2017new}. Machine error has been estimated by running Newton-Raphson iterations until no further convergence is detected and using Kr\"oger's approximant as initial guess.}%
\label{Accuracy_energy.pdf}%
\end{center}
\end{figure}

To have an energy-related error in all the domain, we have computed the integral

\begin{equation}
\mathcal{I}(x)=\int_{0}^{x}\left(  1-\xi\right)  \mathcal{L}^{-1}\left(  \xi\right)  d\xi
\label{int2}
\end{equation}
and report the relative error of this integral in \% against the analytical solution in Figure \ref{Accuracy_energy.pdf}. For the spline-based procedure, we have used $n=10^4$ subintervals and $x_r=0.943$, i.e. our proposal for the  discretization which results were reported also in Figure \ref{Comparison_spline_different_n.pdf}. It is seen that both figures bring to similar conclusions.   

\section{Conclusions}

The inverse Langevin function is used in the evaluation of the behavior of polymers and in other fields like magnetism, molecular dynamics and biomechanics. For example, its efficient and accurate evaluation is of paramount importance in finite element simulations of polymers using chain-based models.

In this paper we propose a new efficient, simple and highly accurate method to evaluate the inverse Langevin function. The present approach is clearly different from that of previous works. The method is purely computational, in line with contemporary methods to solve physics and engineering problems. The method has a simple implementation in any numerical code
in which the evaluation of the inverse Langevin function is needed.
Obviously, the method has more difficult applicability in analytical works, where computers do not play a relevant role and compact expressions are often desired. In these cases, analytical approximants are more adequate.

A study of the accuracy of the method has been also performed, showing  a customizable accuracy which can be taken to almost machine precision. At the same time, the efficiency in the evaluation of the function is in the same order as that  of the most efficient approximants proposed to date and this efficiency is independent of the obtained accuracy. In this sense, the only practical drawback of the method is the use of more RAM memory than traditional approximants. Higher accuracy simply implies more RAM\ memory used, but this memory usage is in any case negligible when compared to the memory available today in any electronic device. Fully commented Matlab codes to reproduce the results given in this paper are attached in the appendices.

\section*{Acknowledgements}
Partial financial support for this work has been given by grant DPI2015-69801-R from the Agencia Estatal de Investigaci\'on of Spain. We thank the reviewers for their comments; they have substantially improved the manuscript.

\section*{Appendix A. Matlab code to generate spline coefficients}
The following Matlab function generates the coefficients of the spline-based procedure. We include several options. This function needs to be run only once in your computer with your choice of custom parameters, since coefficients are saved in Matlab file \tt InverseLangevinCoefs.mat \rm for further use.
\begin{Verbatim}[fontsize=\small]
function InverseLangevinSpline
% InverseLangevinSpline
% Function to create the splines for the Inverse Langevin function
% and save them in Matlab's workspace file 'InverseLangevinCoefs.mat'
%
% Coefs = Coefficients of splines. Last row contains spline breaks
%         For first column, Coef(1:5,1) data are:
%         nsp = number of splines (Coefs has nsp+1 columns)
%         dx  = size of each spline piece 
%         xir = starting point of the rational function
%         a   = coefficient of rational function (a*x+b)/(1-x^2) 
%         b   = coefficient of rational function (a*x+b)/(1-x^2)

%% SELECTION: You can make here your discretization and cut-off choices
%   Some possibilities that have different memory needs and accuracy:     
    %npoints = 11;    xir = 0.860; % => yir(0.860) = 7.142793372503663
    %npoints = 101;   xir = 0.890; % => yir(0.890) = 9.090906992051151
    %npoints = 1001;  xir = 0.925; % => yir(0.925) = 13.333333332400674
    %npoints = 10001; xir = 0.943; % => yir(0.943) = 17.543859649122449
    npoints = 100001; xir = 0.980; % (this gives near machine precision)
    % --- You can change npoints and xir above to any desired value.
    f = @(y) xir-coth(y)+1/y; yir = fzero(f,1/xir); % Computes (xir,yir)
    dyir = yir^2/(1+yir^2-yir^2*(coth(yir))^2);     % dy/dx @ (xir,yir)
%% STEP 1: Discretization of the Langevin function 
    y1 = linspace(0,yir*2,npoints*2-1);    %"*2" avoids border effects
    x1 = coth(y1)-1./y1; x1(1)=0;          % Langevin function values
%% STEPS 2 and 3: Interpolation of (xi,yi) through cubic splines
    sp1=csape(x1,y1,'second');         
%% STEP 4: Computationally efficient discretization
    % Equispacement of abscissae
    x2=linspace(0,xir,npoints); dx = x2(2)-x2(1);
    % Langevin function calculated by the splines 'sp1' 
    y2=fnval(sp1,x2);
    % New set of splines. Prescribe known first derivatives at ends
    sp2=csape(x2,[3,y2,dyir],[1,1]);
    % Number of last spline
    nsp = npoints - 1;
%% STEP 5: Rational function between x=xir and x=1. Compute coefficients
    % Continuity at x=xir
    btemp = yir*(1-xir^2);           % this is b in (a*(x-xir)+b)/(1-x^2)
    % Continuity of the first derivative at x=xir
    Coef(4,1) =-2*btemp*xir/(1-xir^2) +...
         dyir*(1-xir^2);             % this is a in (a*x+b)/(1-x^2) 
    Coef(5,1) = btemp-Coef(4,1)*xir; % this is b in (a*x+b)/(1-x^2)
    % Store some data in first column (to have just one item saved)
    Coef(1,1) = nsp; Coef(2,1) = dx; Coef(3,1) = xir; 
%% SAVE: Retrieve spline coefficients in [0,xir] and save them in file
    coef_1=fnbrk(sp2,'coefs');       % coefficients in pp-form
    % Coefficients between x=0 and x=xir
    Coef(1:4,2:nsp+1)= coef_1(1:nsp,1:4)'; % transposed for efficiency
    Coef(5,2:nsp+1)  = x2(1:nsp);    % save spline breaks in Coef
    % Save final coefficients in a file for future "load" in Matlab
    save('InverseLangevinCoefs','Coef');
return
end
\end{Verbatim}

\section*{Appendix B. Matlab code to compare approximants}
In this appendix we include a Matlab script to compare accuracy and efficiency of approximants. This script generates figures similar to Figures \ref{Accuracy1e5pieces.pdf} and \ref{CPU_vf_long_time2.pdf}. Figures \ref{Comparison_spline_different_n.pdf} and \ref{Accuracy_energy.pdf} may be obtained by straightforward modifications. 
\begin{Verbatim}[fontsize=\small]
%% Construction of the Inverse Langevin Function spline
    % This section needs to be run only once in your computer
    % Creates Coefs and saves them in 'InverseLangevinCoefs.mat'
    close all; clear all; InverseLangevinSpline;
%% Load coefficients of spline (they can be embeded in compiled code)
    load('InverseLangevinCoefs');
    nsp = Coef(1,1); dx = Coef(2,1); xir  = Coef(3,1);
    a   = Coef(4,1); b  = Coef(5,1); nsp1 = nsp+1;
%% ======  EVALUATION OF ERRORS AND CPU TIME OF APPROXIMANTS ======= 
    ntimes = 1000000; y_max = 1000; % number of times and max. value
    y_exact = (1e-2 + rand(ntimes,1)*y_max)'; % 1e-2 avoids "zero"
    x = coth(y_exact) - 1./y_exact;           % Langevin function
%% Spline-Based approximation (present proposal)
    t=cputime; 
    for i=ntimes:-1:1    
        s = fix(x(i)/dx)+2; % Spline index 
        if s < nsp1         % Use cubic spline polynomial #(s-1): 
            xinc = x(i)-Coef(5,s); xinc2 = xinc^2; xinc3 = xinc2*xinc; 
            ySpline(i) = Coef(1,s)*xinc3 + Coef(2,s)*xinc2 +...
                            Coef(3,s)*xinc + Coef(4,s);  
        else                % Use rational function near asymptote:
            ySpline(i) = (a*x(i)+ b)/(1-x(i)^2);
        end
    end
    Spline_cpu = cputime-t, 
%% Kroger 2015. Int. J. Non-Newtonian Fluid Mechanics 223:77-87
    t=cputime;
    for i=ntimes:-1:1    
        x2 = x(i)^2; x4 = x2^2;  
        yKroger(i)=(3*x(i)-x(i)/5*(6*x2+x4-2*x2*x4))/(1-x2);
    end
    Kroger_cpu = cputime-t,
%% Petrosyan 2017. Rheologica Acta 56:21-26.
    t=cputime;
    for i=ntimes:-1:1   
        x2 = x(i)^2; 
        yPetrosyan(i)=3*x(i)+(x2)/5*(sin(7*x(i)/2))+x(i)*x2/(1-x(i));
    end
    Petrosyan_cpu = cputime-t,
%% Nguessong et al 2014. Rheologica Acta 53:585-591. 
    t=cputime;
    for i=ntimes:-1:1     
        x2 = x(i)^2;
        yNguessong(i)=x(i)*(3-x2)/(1-x2)-0.488*x(i)^3.243+...
                 3.311*(x(i)^4.789)*(x(i)-0.76)*(x(i)-1);
    end
    Nguessong_cpu = cputime-t,
%% Marchi & Arruda 2015. Rheologica Acta 12:887-902.
    t=cputime;
    for i=ntimes:-1:1   
        yMarchi(i)=x(i)*(3-0.631531*x(i)-0.578498*x(i)^2)/...
            ((x(i)-1)*(-1-0.789957*x(i)))-0.44692*x(i)^4.294733-...
            (11.08867*x(i)^11.60749)*(x(i)-1.004823)*(x(i)-1.022831);
    end
    Marchi_cpu = cputime-t,   
%% Jedynak 2017. J. Non-Newtonian Fluid Mechanics 249:8-25. 
    t=cputime; 
    for i=ntimes:-1:1       
        x2 = x(i)^2; x4 = x2^2;
        yJedynak(i)=x(i)*(3-1.00651*x2-0.962251*x4+...
            1.47353*x4*x2-0.48953*x4*x4)/...
            ((1-x(i))*(1+1.01524*x(i)));
    end
    Jedynak_cpu = cputime-t, 
%% Newton-Raphson: Gives machine precision. Needs one approximant.
    disp('* Newton-Raphson: Please, be patient, this will take time *');
    t=cputime;
    yNewton = yKroger;  % use Kroger's approximant for a starting guess
    nit = 0; big = 10^8; Big = 10^10; 
    for i=ntimes:-1:1
        residual = big; residual_old = Big;      % Initial big numbers
        while(abs(residual) < abs(residual_old)) % Until no further conv.
            residual_old = residual; 
            residual = coth(yNewton(i)) - 1/yNewton(i) - x(i);
            nit = nit + 1; y2 = yNewton(i)^2;
            dyNewton = y2/(1+y2-y2*coth(yNewton(i))^2);
            yNewton(i) = yNewton(i) - dyNewton * residual;
        end,          % At this point we reached best possible precision
    end,
    Newton_cpu = cputime - t,
    Mean_number_of_iterations_per_point_over_Kroger = nit/ntimes,
%% Plot accuracy results
    figure; hold on;
    hS = plot(x,abs(ySpline-y_exact)./y_exact*100,'*','color',[0.5,0.5,0]);
    hK = plot(x,abs(yKroger-y_exact)./y_exact*100,'ko');   
    hP = plot(x,abs(yPetrosyan-y_exact)./y_exact*100,'bs'); 
    hN = plot(x,abs(yNguessong-y_exact)./y_exact*100,'c^'); 
    hM = plot(x,abs(yMarchi-y_exact)./y_exact*100,'mp'); 
    hJ = plot(x,abs(yJedynak-y_exact)./y_exact*100,'rv'); 
    hE = plot(x,abs(yNewton-y_exact)./y_exact*100,'g.'); 
    lK = 'Kroger'; lP = 'Petrosyan'; lN = 'Nguessong et al'; 
    lM = 'Marchi-Arruda'; lJ = 'Jedynak'; lS = 'Spline-based';
    lE = 'Machine error (N-R)'; leg = {lS;lK;lP;lN;lM;lJ;lE};   % Legend
    hleg = legend([hS,hK,hP,hN,hM,hJ,hE],leg);
    set(hleg,'units','normalized','position',[0.4,0.28,0.2,0.2]); 
    xlabel('x-variable (Langevin function)'); 
    ylabel('Inverse Langevin fnct. (force) % relative error (Log scale)');
    title('Relative error (%) of approx. to the inverse Langevin funct.');
    set(gca,'YScale','log'); set(gca,'ylim',[1e-14,1]);
%% Plot CPU timing results
    figure;
    barh([1:6],[Spline_cpu,Kroger_cpu,Petrosyan_cpu,Nguessong_cpu,...
             Marchi_cpu,Jedynak_cpu]);  hold on;
    plot([Spline_cpu,Spline_cpu],[0,7],'r','linewidth',2);    
    set(gca,'YTickLabel',leg);
    xlabel('CPU time of approximants [seconds]');
    title(['CPU time of ',num2str(ntimes),' random evaluations from',...
           ' y ~ 0 to y = ',num2str(y_max)]);

\end{Verbatim}

\section*{References}

\bibliography{Benitez_Montans_JNNFM2018_R1}

\end{document}